\documentclass{article}
\usepackage{times}
\usepackage{graphicx}

\begin{document}

\title{``Soft bang'' instead of ``big bang'':\\
  model of an inflationary universe without singularities and with
  eternal physical past time}  
\author{E.~Rebhan\\ Institut f{\"u}r Theoretische Physik,
  Heinrich--Heine--Universit{\"a}t,\\
  D--40225 D{\"u}sseldorf, Germany}

\date{Received 28 August 1998 / Accepted 14 October 1999}


\maketitle

\begin{abstract}
  The solution for an inflationary universe without singularities is
  derived from the Einstein-Lema{\^\i}tre equations.  The present
  state of the universe evolved from a steady state solution for a
  tiny, but classical micro-universe with large cosmological constant
  or large equivalent vacuum energy density and with an equal energy
  density of radiation and/or some kind of relativistic primordial
  matter in the infinite past. An instability of this state outside
  the quantum regime caused a ``soft bang'' by triggering an expansion
  that smoothly started with zero expansion rate, continuously
  increased, culminated in an exponentially inflating phase and ended
  through a phase transition, the further evolution being a
  Friedmann-Lema{\^\i}tre evolution as in big bang models.  As a
  necessary implication of the model the universe must be closed.  All
  other parameters of the model are very similar to those of big bang
  models and comply with observational constraints.

\end{abstract}



\section{Introduction}

In classical big bang models the universe starts at infinite density
and infinite temperature with infinite expansion velocity and
expansion rate. No reason can be given why it expands so rapidly, the
conditions at the big bang enter the models as unexplained initial
conditions. In the case of an open universe these conditions must even
prevail throughout the infinite space which the universe occupied
already at the very beginning. Since for matter densities above the
Planck-density~$\varrho_{\mathrm{Pl}}$ quantum-gravitational effects
become important, the validity of classical big bang models is
restricted to densities below $\varrho_{\mathrm{Pl}}$ and times after
the Plank time $t_{\mathrm{Pl}}$ at which this density is reached.
Theories describing a quantum-evolution of the universe have been
presented by several authors: de Witt (\cite{witt}), Wheeler
(\cite{wheeler}), Hartle \& Hawking (\cite{hartle}) and others. In
general, singularities cannot be avoided in these models either.
However, when it is assumed that at very early times there are no
particles but only vacuum fields in the universe, then nonsingular
solutions also exist, having a similar time evolution as the model
considered in this paper (Starobinsky \cite{starobinsky}).

A serious problem of former big bang models is the enormously large
value of about $10^{30}$ times the Planck-length $l_{\mathrm{Pl}}$
obtained for the cosmic scale factor at the Plank time, spanning a
range that is not causally connected but nevertheless must be the
source of the fantastically isotropic microwave background radiation
observed. This problem has been overcome by incorporating into the big
bang models a phase of inflationary expansion (Guth \cite{guth}, Linde
\cite{linde} and Albrecht \& Steinhardt \cite{albrecht}), according to
which within an extremely short time a causally connected region with
the extension of a Planck-length is inflated to the much larger domain
obtained from the classical Friedmann-Lema{\^\i}tre evolution.
Exponential expansion of an inflationary phase can be obtained for a
cosmic substrate with positive energy density and negative pressure
(Gliner \cite{gliner}), and can be explained by the presence of a
scalar Higgs field as assumed in the Grand Unified Theories (Georgi \&
Glashow \cite{georgi}), other scalar fields, tensor fields or other
mechanisms (see Overduin \& Cooperstock \cite{overduin} for a list
of references). The existence of a field of this kind is equivalent to
the existence of a - dynamical - cosmological constant, and in the
early stages of the universe, the value of the latter must be
extremely high, $\Lambda\approx 10^{53}/\mathrm{m}^{2}$, if the above
mentioned difficulty should be overcome.

Thus, there are strong arguments supporting the existence of a large
cosmological constant $\Lambda$ in the very early universe even in the
big bang models, classical or combined quantum/classical, and when
speaking about big bang models, it will further be assumed that an
inflationary phase due to the presence of a large $\Lambda$ is
included.  When this necessity is accepted, then cosmological models
become possible that avoid the singularities of classical big bang
models and even avoid the necessity to invoke a theory of quantum
gravity or rather precursors of such a theory which does not yet
exist.

A singularity-free model was suggested by Israelit \& Rosen
(\cite{israelit}) (called IR-model in the following) according to
which the (closed) universe was set into existence as a tiny bubble in
a homogeneous and isotropic quantum state with a pre-material vacuum
energy density corresponding to $\varrho_{\mathrm{v}} =
\varrho_{\mathrm{Pl}}=5.2\cdot 10^{96}\,\mathrm{kg/m}^{3}$, with
matter and/or radiation density $\varrho_{\mathrm{m}}=0$ and with the
diameter of a Planck length as an initial condition. In this model it
is assumed that, because it is in the quantum regime, this state can
prevail for some time. At the moment when it traverses the barrier to
classical behavior it experiences an accelerated expansion described
by the expanding branch of a de Sitter solution. For the later
evolution a standard model solution with $\Lambda=0$,
$\Omega_{0}=1.16$ and $H_{0}=46.5
\,\mathrm{km}\,\mathrm{s}^{-1}\mathrm{Mpc}^{-1}$ is assumed. The
transition from de Sitter inflation into this evolution is performed
by a phase transition (Kirzhnits \cite{kirzhnits}, Kirzhnits \&
Linde~\cite{kirzhnitslinde} and Albrecht \& Steinhardt
\cite{albrecht}), transforming vacuum energy density into ordinary
matter as in the inflation scenarios of big bang models. In order to
get a smooth connection to a reasonable later evolution, this
transition must occur at the very early time $t\approx 100
\,t_{\mathrm{PL}}$ when the density of the standard model solution is
$\varrho_{\mathrm{m}}\approx \varrho_{\mathrm{Pl}}\cdot
(t_{\mathrm{PL}}/t)^{3/2} \approx 10^{-3}\varrho_{\mathrm{Pl}}$. This
is still far above the density of $\approx 10^{86}\,\mathrm{kg/m}^{3}$
that prevails in this model at the temperature corresponding to
$10^{16}\,\mathrm{GeV}$ at which GUTs with simple groups give rise to
magnetic monopoles of mass $m_{M}\approx 10^{16}\,\mathrm{GeV/c^{2}}$
('t Hooft \cite{hooft}, Polyakov \cite{polyakov} and Zeldovich \&
Novikov \cite{zeldovichnovikov}). Since inflation is already over
at this stage, no possibility exists in this model to dilute the heavy
monopoles to such a low concentration that an expansion is possible
until the present, and that the detection of magnetic monopoles is
most unlikely as expressed by the fact they have not yet been
observed.

Blome \& Priester (\cite{blome}) proposed a ``big bounce model''
(called BP-model in the following) in which the universe is closed,
exists for an infinite time and first contracts from an infinite size.
After it has passed through a minimal radius larger than
$l_{\mathrm{Pl}}$ it starts re-expanding.  In the contraction phase
the universe is in a primordial state of a quantum vacuum with finite
energy density and negative pressure $p$. The ``big bounce'' at the
minimal radius is supposed to trigger a phase transition by which
ordinary matter is created, the vacuum energy density is reduced and
the transition into a Friedmann-Lema{\^\i}tre evolution is achieved as
in the inflation scenarios of big bang models. The parameters of the
model are adjusted in a way such that the above mentioned monopole
condition is satisfied.

The IR-model was further developed by Starkovich \& Cooperstock
(\cite{starkovich}) and by Bayin et al. (\cite{bayin}).  The
pre-material vacuum energy density of the IR-model is attributed to a
scalar $\Phi$ field obeying a covariant Klein-Gordon equation that is
extended by an additional coupling term to the gravitational field and
a potential of the field.  Furthermore, an equation of state of the
form $p=(\gamma -1)\varrho_{\mathrm{v}}c^{2}$ for the pressure $p$ is
assumed, for the entropy density the equation of state
$s=\gamma\varrho/T$ is derived, and an adiabatic evolution according
to \mbox{$sS^{3}$= const} is found to comply with the other equations.
Three different epochs in the evolution of the universe are assumed
for each of which $\gamma$ is treated as a constant: first an
inflationary era with very small $\gamma$, second a radiation era with
$\gamma=4/3$, and third a matter era with $\gamma=1$. The starting and
end points of the different eras are determined by critical or
limiting values of physical quantities like the Planck density or the
Planck temperature, the idea of taking these as limiting values for
cosmological models being adopted from Markov (\cite{markov}).  As in
the IR-model the (closed) universe starts at the Planck density
$\varrho_{\mathrm{Pl}}$ with $S=l_{\mathrm{Pl}}$, and in addition to
the IR-model the condition $\dot{S}=0$ is imposed as initial
condition, thus providing a start without singularity. (The model
starts at the minimum radius of a big bounce model and, in principle,
could be extended further into the past by coupling it to a
contraction phase like in the BP-model.)  In the inflationary phase,
due to the small value of $\gamma$ the density $\varrho_{\mathrm{v}}$
decreases only very slowly whence $T\sim\varrho_{\mathrm{v}}S^{3}$
increases, and the inflation is ended when $T$ reaches the Planck
temperature~$T_{\mathrm{Pl}}$.  This way no fine tuning is needed for
the adjustment to the following radiation dominated era, and in
addition, no re-heating is needed because, differently from usual
inflation models, the universe enters the radiation era with the
appropriate temperature.

Like big bang models with inflation, the model presented in this paper
requires some primordial relativistic matter and/or radiation in
addition to a high energy density of some primordial quantum field. It
can even avoid singularities that are still present in the BP-model,
i.e. infinite extension of the closed universe and infinite velocity
of contraction in the far past, $\dot{S}=-\infty$, and it also avoids
the problem of missing monopole dilution.

\section{The ``soft bang'' model}
\label{sec:softb}

The equations for a homogeneous and isotropic universe to be used are
the well known Lema{\^\i}tre equations
\begin{equation}\label{1}
  \dot{S}^{2} 
  = \frac{8\pi G}{3} \varrho S^{2}
  + \frac{\Lambda c^{2}}{3}\,S^{2} - k c^{2} \,,
\end{equation}
\begin{equation}\label{2}
  \ddot{S} = - \frac{4\pi G}{3} \biggl(\varrho + 
  \frac{3 p}{c^{2}}-\frac{\Lambda c^{2}}{4\pi G}\biggr) \,S
\end{equation}
($G=$ gravity constant, $\Lambda=$ cosmological constant, $c=$ speed
of light, $k=-1,0,1$, $S=$~cosmic scale factor, \mbox{$\varrho=$ mass}
density and $p=$ pressure) following from Einsteins
field equations. Eq.~(\ref{2}) is a consequence of Eq.  (\ref{1}) for
$\dot{S}\ne 0$ when the energy equation
\begin{equation}\label{3}                                               
\frac{d}{dt}(\varrho S^{3}) = - \frac{p}{c^{2}} \,\frac{d}{dt} S^{3}
\end{equation}
is employed.

Concerning the cosmological constant, as in the usual scenarios of
cosmic inflation it is assumed that it can be explained and assumes a
very large value due to the presence of some primordial vacuum field
which at early times and very high temperatures was in a ground state
of very high energy density. There are two possible ways
to incorporate this assumption into Eqs. (\ref{1})--(\ref{2}):\\
1.  One can set $\Lambda=0$ and replace $\varrho\to
\varrho_{\mathrm{m}}+\varrho_{\mathrm{v}}$, where
$\varrho_{\mathrm{v}}$ is the mass density of a quantum field
corresponding to its energy density, and $\varrho_{\mathrm{m}}$ is the
mass density of matter plus radiation. In this case the equations must
be supplemented with the ansatz
$p_{\mathrm{v}}=-\varrho_{\mathrm{v}}c^{2}$ for a negative pressure of
the vacuum (Starobinsky \cite{starobinsky} and Zeldovich
\cite{zeldovich}), and it must be assumed that $p_{\mathrm{v}}$ adds
to the pressure $p_{\mathrm{m}}$ of radiation and matter, yielding the
total pressure $p=p_{\mathrm{m}}+ p_{\mathrm{v}}$.\\ 2. Equivalently
one can set $\varrho=\varrho_{\mathrm{m}}$, $p=p_{\mathrm{m}}$ and
replace $\Lambda$ by $(8\pi G/c^{2}) \varrho_{\mathrm{v}}$.  If in
addition the equation of state for relativistic matter and radiation,
$p_{\mathrm{m}}= \varrho_{\mathrm{m}}c^{2}/3$, is employed, then from
(\ref{3}) both representations yield
\begin{equation}\label{4}
  \frac{\dot{\varrho}_{\mathrm{m}}}{\varrho_{\mathrm{m}}} = 
      \frac{4\dot{S}}{S}\qquad\mbox{or}\qquad
      \varrho_{\mathrm{m}}S^{4}=\varrho_{*}S^{4}_{*}
\end{equation}
where $\varrho_{*}$ and $S_{*}$ are constant
values of $\varrho_{\mathrm{m}}$ and $S$ to be specified later, and from
(\ref{1})--(\ref{2}) both representations yield
\begin{eqnarray}
  \dot{S}^{2}(t) &=& \frac{8\pi G}{3}\,
  (\varrho_{\mathrm{m}} + \varrho_{\mathrm{v}}) S^{2} - k c^{2}\,, \label{5}\\
 \ddot{S}(t) &=&- \frac{8\pi G}{3}\big(\varrho_{\mathrm{m}} 
 - \varrho_{\mathrm{v}}\big)\,S\,.\label{6}
\end{eqnarray}

Now the first step is to look for steady state solutions. For these
$\dot{S}=\ddot{S}=0$ is required, and from $\ddot{S}=0$ and (\ref{6})
the condition
\begin{equation}
  \label{7}
  \varrho_{\mathrm{m}}=\varrho_{\mathrm{v}}=:\varrho_{*}\,.
\end{equation}
is obtained. It implies that in the static initial state from which
the universe is supposed to evolve, there must be an equipartition
between the energy of the vacuum and the energy of relativistic
matter and/or radiation.  Using this result, from the second
requirement $\dot{S}=0$ and from (\ref{6}) the conditions $k=1>0$ and
\begin{equation}
  \label{8}
  S=S_{*}:=c\,\sqrt{\frac{3}{16\pi G \varrho_{*}}}
\end{equation}
are obtained. It must, of course, be expected that, like Einstein's
static universe, this static solution will be unstable. However, it is
just this property which opens the possibility that the present
universe has evolved from it through an instability.

The instability of the static solution (\ref{8}) follows from the
existence of a dynamic solution that asymptotically converges towards
it as $t\to -\infty$. To prove this and to derive the unstable
solution it suffices to solve Eq. (\ref{5}) because for $\dot{S}\ne 0$
Eq. (\ref{6}) will then be satisfied automatically. In order to
satisfy the condition imposed on the asymptotic behavior,
$\varrho_{\mathrm{v}}=\varrho_{*}$ and (\ref{8}) must be
inserted in Eqs. (\ref{4}) and (\ref{5}).  However, it is more
convenient to use (\ref{8}) and replace $\varrho_{*}$ with 
\begin{displaymath}
  \varrho_{*}=\frac{3c^{2}}{16\pi G S^{2}_{*}}\,,
\end{displaymath}
and from (\ref{5}) one thus obtains 
\begin{equation}  \label{9}
  \dot{S}^{2}=\frac{c^{2}}{2}
  \left(\frac{S^{2}_{*}}{S^{2}}+\frac{S^{2}}{S^{2}_{*}}-2\right)
  =\frac{c^{2}}{2}\left(\frac{S}{S_{*}}-\frac{S_{*}}{S}\right)^{2}
\end{equation}
or
\begin{equation}  \label{10}
  \dot{S}=\pm\frac{c}{\sqrt{2}}\left(\frac{S}{S_{*}}-\frac{S_{*}}{S}\right)\,.
\end{equation}
For the equation with the plus sign one easily finds the solution
\begin{equation}
  \label{11}
  S=S_{*}\left(1+e^{\sqrt{2}\,c(t-t_{0})/S_{*}}\right)^{1/2}
\end{equation}
with $t_{0}$ being an integration constant. (The solution for the
minus sign, obtained from this solution by replacing $t-t_{0}\to
t_{0}-t$, has $\ddot{S}<0$ instead of $\ddot{S}>0$ and is of no
interest here because it describes a contracting universe.)  For the
solution (\ref{11}) the universe has already existed for an infinite
time and was separated from the static solution (\ref{8}) through an
instability in the infinite past. First it starts expanding very
slowly, the expansion velocity $\dot{S}$ becoming larger and larger
with time until the exponential term becomes dominant and the
expansion exponentially inflating according to
\begin{displaymath}
  S=S_{*}e^{c(t-t_{0})/(\sqrt{2}\,S_{*})}
\end{displaymath}
just as after a big bang. During inflation the matter is getting
extremely diluted and, as a consequence of this, cooled down as well.
Therefore, in principle it would be necessary to replace (\ref{4}b) at
a certain stage by the equation
$\varrho_{\mathrm{m}}S^{3}=\tilde{\varrho}\tilde{S}^{3}$ valid for
cold matter. However, this is not necessary because at this stage the
matter density $\varrho_{\mathrm{m}}$ is much smaller than the vacuum
density $\varrho_{\mathrm{v}}$ and can thus be neglected.

It must be assumed that through the process of inflation a phase
transition is triggered, transforming energy of the vacuum field into
energy of matter as in the usual inflation models. From the
simplifying assumption that this transition occurs instantaneously at
the scale-factor $S_{1}=S_{\mathrm{F}}(t_{1})$ of a
Friedmann-Lema{\^\i}tre solution for regular matter extending until
today, and from (\ref{11}), the condition
\begin{displaymath}
    S_{1}=S_{*}\left(1+e^{\sqrt{2}\,c(t_{1}-t_{0})/S_{*}}\right)^{1/2}\,,
\end{displaymath}
is obtained, which is satisfied by the choice
\begin{displaymath}
  t_{0}=t_{1}-\frac{S_{*}}{\sqrt{2}\,c}\,
  \ln \left[\left(\frac{S_{1}}{S_{*}}\right)^{2}-1\right]
\end{displaymath}
of the integration constant $t_{0}$.

\begin{figure}[tb]
  (a)\includegraphics[width=0.9\columnwidth]{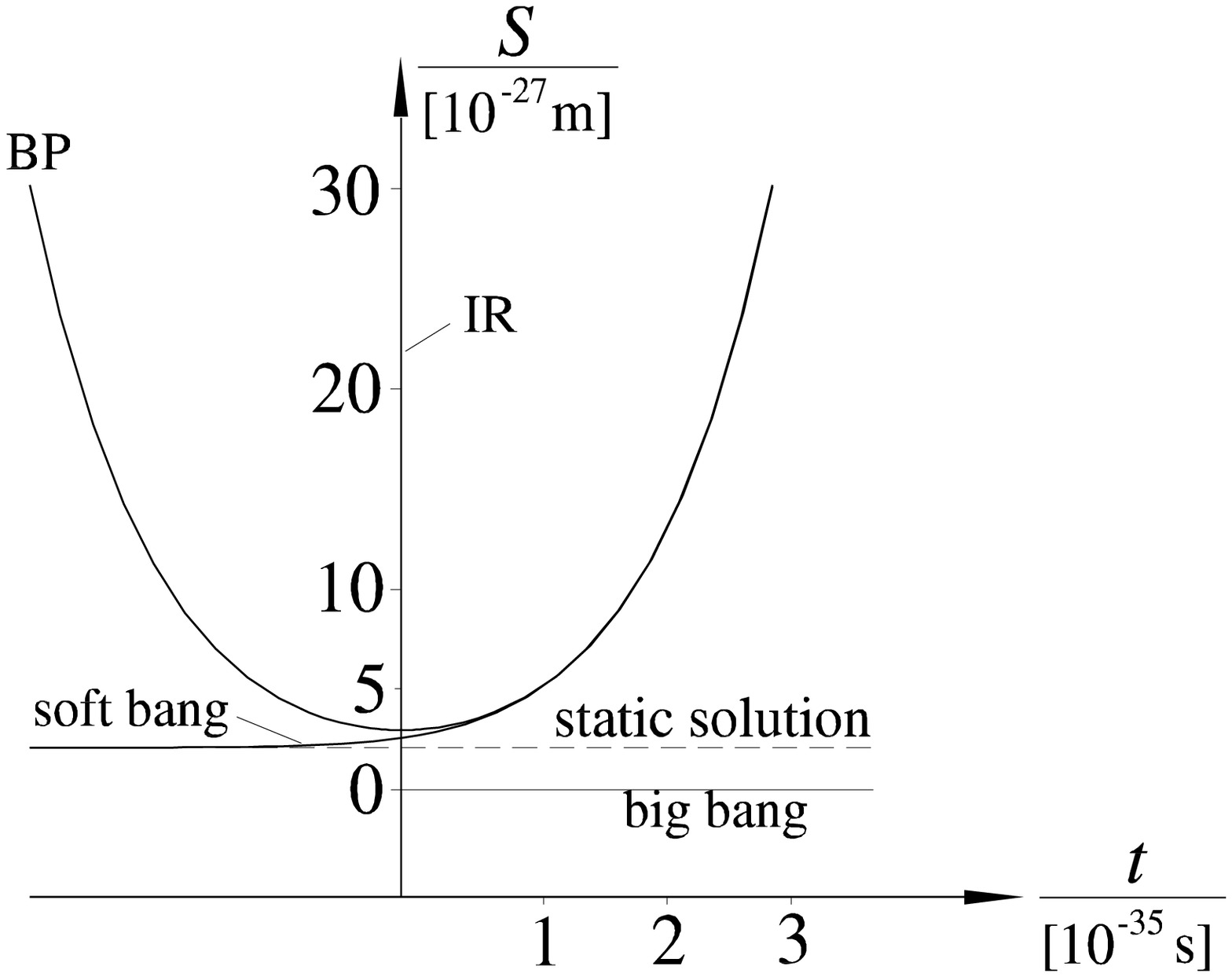}\\
  (b)\includegraphics[width=0.7\columnwidth]{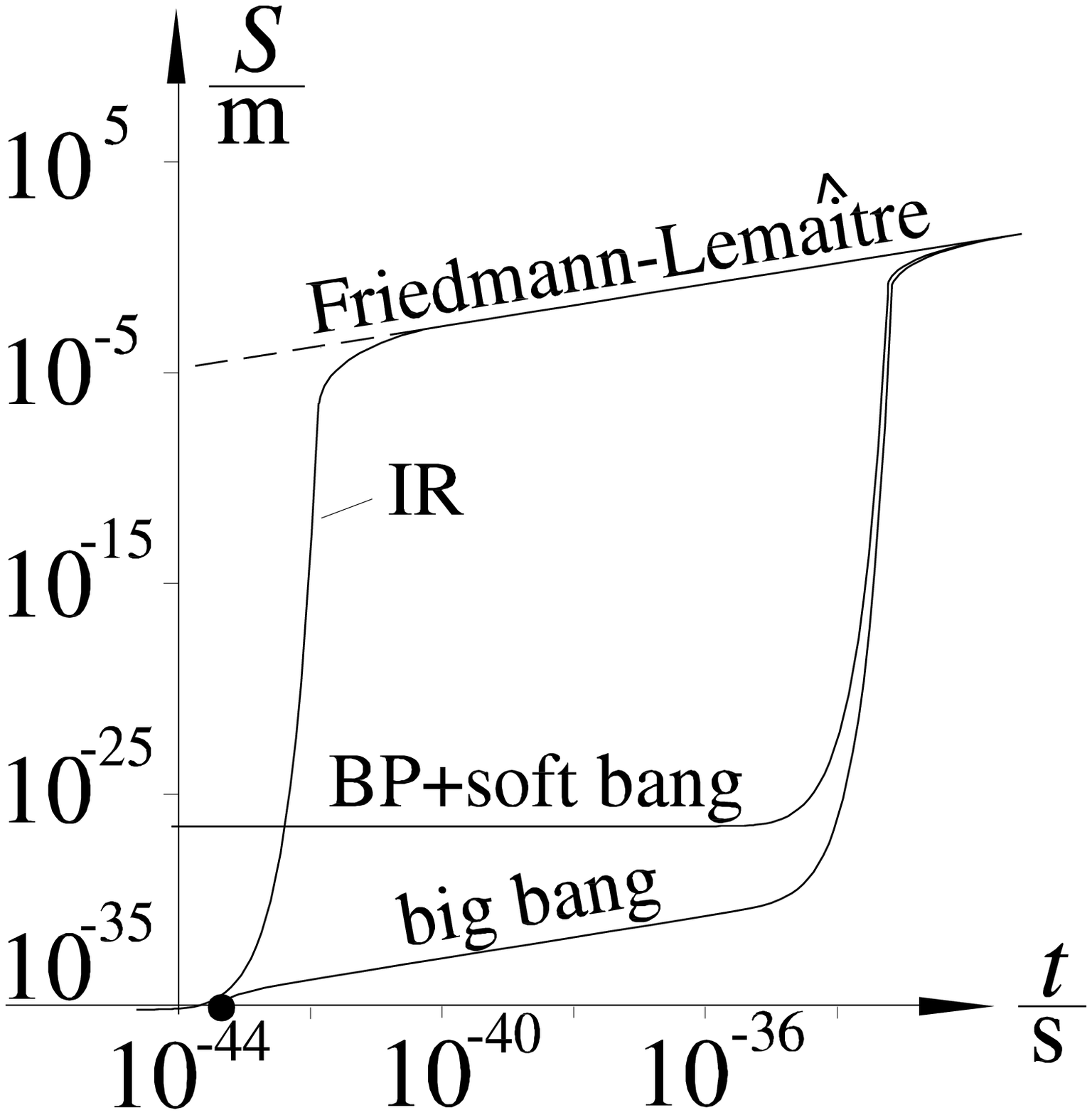}\\
  \caption{\label{pic:1}
    Evolution of $S(t)$ for the model of this paper in comparison with
    the IR-model, the BP-model and a big bang model (a) for times
    before and after the Planck-time in linear time scale, and (b) for
    $t\ge t_{\mathrm{Pl}}$ in logarithmic time scale. In the linear
    time scale according to the IR-model $S(t)$ increases so rapidly
    that it coincides with the $S$-axis, while for the big bang model
    $S(t)$ is still so small that it coincides with the $t$-axis. For
    $t\ge t_{\mathrm{Pl}}$ in logarithmic time scale the present model
    and the big bounce model become indistinguishable.  After
    inflation all models merge into the same Friedmann-Lema{\^\i}tre
    evolution.}
\end{figure}

Since in the phase of inflation, according to (\ref{4}), radiation and
relativistic matter of the joint density $\varrho_{\mathrm{m}}$ are
becoming extremely diluted, it must be assumed that energy of the
vacuum field with density $\varrho_{\mathrm{v}}c^{2}$ is transformed
into energy of regular matter from which the present matter of the
universe derives.  Therefore, the time $t_{1}$ must be before the
creation of quarks but late enough that no remarkable density of
magnetic monopoles was able to develop, which would happen for
$T(t_{1})\ge 10^{29}\,\mathrm{K}$.  With the choice
$\varrho_{*}=2\cdot 10^{79} \,\mathrm{kg}/\mathrm{m}^{3}$
corresponding to $S_{*}\approx 3\cdot 10^{-27}\,\mathrm{m}$, according
to (\ref{8}) the intersection of the solution (\ref{11}) with a
Friedmann-Lema{\^\i}tre solution $S_{\mathrm{F}}(t)$ (obtainable from
(\ref{12})) occurs at the time $t_{1}\approx 10^{-33}\,\mathrm{s}$
with $T(t_{1})\approx 4\cdot 10^{26}\,\mathrm{K}$, and both conditions
are met. The radius $S_{*}\approx 3\cdot 10^{-27}\,\mathrm{m}$ from
which the universe started according to the present model is well
above the Planck length by a factor of $\,\approx 10^{8}$.  Thus the
present model is far out of the regime where quantum gravity has to be
employed.  Fig.~\ref{pic:1} shows the time evolution of $S(t)$ for the
present model ($\varrho_{*}= 2\cdot 10^{79}
\,\mathrm{kg}/\mathrm{m}^{3}$), for a big bang model with inflation
($\varrho_{\mathrm{v}}= 2\cdot 10^{79} \,\mathrm{kg}/\mathrm{m}^{3}$
and $\varrho_{\mathrm{m}}=\varrho_{\mathrm{Pl}}= 5\cdot 10^{96}
\,\mathrm{kg}/\mathrm{m}^{3}$), for the BP-model
($\varrho_{\mathrm{v}}= 2\cdot 10^{79} \,\mathrm{kg}/\mathrm{m}^{3}$),
and finally for the IR-model
($\varrho_{\mathrm{v}}=\varrho_{\mathrm{Pl}}= 5\cdot 10^{96}
\,\mathrm{kg}/\mathrm{m}^{3}$).

A question of considerable interest is how and why the phase
transition from inflation to ordinary Friedmann-Lema{\^\i}tre
expansion is triggered.  A dynamical evolution, for example some
instability of the vacuum field, must be supposed, and it must be
assumed that it was not present in the infinite time before inflation.
Qualitatively it may be expected that, as in the inflation scenarios
of big bang models, extreme temperatures before the inflation give
rise to thermal fluctuations that change the thermally averaged
potential of a quantum field in such a way that it has a large
positive minimum value. The phase transition can then be attributed to
a decrease of this minimum caused by the rapid cooling through
inflationary expansion.

During the phase transition, the extremely low values of density and
temperature to which the primordial matter and/or radiation have been
brought down through inflation must be restored to the high values
that are required as initial values for the Friedmann-Lema{\^\i}tre
evolution finally leading to the present state of the
universe. For this process several possibilities exist:\\
1. The primordial matter remains as cold and diluted as it came out of
the preceding inflation phase and is cooled down and diluted still
further during the following Friedmann-Lema{\^\i}tre evolution. In
this case the vacuum energy density $\varrho_{\mathrm{v}}$ must be
completely converted into the density of hot matter and radiation, and
matter of a kind completely different from the primordial matter that
existed for $t\to -\infty$ could be created. As a consequence, in
addition to the matter that developed during and after the phase
transition there could be a second component of quite different
matter, although diluted and cooled down extremely.\\
2. The vacuum energy could be completely used up for re-heating the
primordial matter, at the same time increasing its density
due to the mass contained in its thermal energy.\\
3. The third possibility consists in a combination of re-heating of
old matter and creation of new matter from vacuum energy.\\
In this paper, no preference to any of these possibilities is given
because the Friedmann-Lema{\^\i}tre evolution following the phase
transition is the
same for all of them.\\

In the following differences between the present model and big bang
models are searched. Before the phase transition the present model
provides much more time than big bang models, and also the ratio
$\varrho_{\mathrm{m}}/\varrho_{\mathrm{v}}$ is quite different,
because in the big bang model the density $\varrho_{\mathrm{m}}$ has
already dropped to $\varrho_{\mathrm{m}}\approx
10^{-32}\varrho_{\mathrm{Pl}}\approx
2.5\cdot10^{-15}\varrho_{\mathrm{v}}$ when it has reached the scale
factor $S_{*}$, while in the present model
$\varrho_{\mathrm{m}}=\varrho_{\mathrm{v}}$ at this stage.  Therefore,
the stability behavior with respect to perturbations that locally
destroy the high symmetry of the cosmological principle may be quite
different.  However, it must be expected that differences arising this
way are washed out by the exponential inflation. Furthermore, in the
process of phase transition, in spite of quite different values of
$\varrho_{\mathrm{m}}$, no differences can be expected to arise,
because in both models $\varrho_{\mathrm{m}}$ is extremely diluted so
that it can be neglected in comparison with $\varrho_{\mathrm{v}}$
which is the same in both models.

After the period of inflation, both models merge into a
Friedmann-Lema{\^\i}tre evolution first described by the Lema{\^\i}tre
equation
\begin{equation}
  \label{12}
  \dot{S}^{2} 
  = \frac{8\pi G}{3} \frac{\varrho_{1} S^{4}_{1}}{S^{2}}
  + \frac{\Lambda c^{2}}{3}\,S^{2} - k c^{2} 
\end{equation}
and later until today by the Friedmann equation
\begin{equation}
  \label{12*}
  \dot{S}^{2} 
  = \frac{8\pi G}{3} \frac{\varrho_{0} S^{3}_{0}}{S}
  + \frac{\Lambda c^{2}}{3}\,S^{2} - k c^{2}\,. 
\end{equation}
A complete discussion of all possible solutions is given in a survey
paper by Felten \& Isaacman (\cite{felten}).

Since during the phase transition most of the vacuum energy is used up
for re-heating and/or creation of matter, after it $\Lambda$ must be
either zero or have an extremely small value in comparison with its
value before the phase transition. In order to obtain a solution that
complies with the matter density presently observed in the universe,
$\Lambda$ must be different from zero.

As long as $S$ is sufficiently small, the $\Lambda$- and the $k$-term
on the right hand side of (\ref{12}) can be neglected and the
evolution of the present model and big bang models is essentially
identical. Thus the possible appearance of differences is restricted
to later times. From the Friedmann equation, for the quantities
\begin{equation}\label{13}                                               
  \Omega_{0} = \frac{\varrho_{\mathrm{m}}(t_{0})}{\varrho_{\mathrm{crit}}(t_{0})} 
  \,, \;\;
  \lambda_{0} = \frac{\varrho_{\Lambda}}{\varrho_{\mathrm{crit}}(t_{0})}
\end{equation}
with
\begin{displaymath}
  \varrho_{\mathrm{crit}}(t) = \frac{3 H^{2}(t)}{8\pi G} 
\end{displaymath}
the condition
\begin{equation}\label{14} 
  \Omega_{0} + \lambda_{0} - 1 =\frac{k c^{2}}{H_{0}{}^{2}S_{0}{}^{2}}
\end{equation}
with $H=$ Hubble parameter can be derived and becomes
\begin{equation}\label{15}
  \Omega_{0} + \lambda_{0}=1+\frac{c^{2}}{H_{0}{}^{2}S_{0}{}^{2}}
\qquad\mbox{or}\qquad
   \Omega_{0} =1 + \frac{k c^{2}}{H_{0}{}^{2}S_{0}{}^{2}}
\end{equation}
for the present model ($k=1$) or the standard model ($k=-1,0,1$ and
$\lambda_{0}=0$) respectively. Furthermore, for the deceleration
parameter $q_{0}=-\ddot{S}_{0}/(H^{2}_{0}S_{0})$ the result
\begin{equation}\label{16}
  q_{0}=\frac{\Omega_{0}}{2}-\lambda_{0}
\end{equation}
is obtained.

In the present model, as in the standard model, the choice
$\lambda_{0}=0$ is possible, and in this case no difference between the
two arises if $k=1$ is chosen in the latter as well. In the standard
model $\Omega_{0} =1$ for $k=0$, and when in the present model
$\lambda_{0}$ is zero and $\Omega_{0}$ is only slightly above $1$,
then the differences between the two models are again
negligible. 

The values $\Omega_{0}\ge 1$ that have to be chosen in the cases
$\lambda_{0}=0$ and $k\ge 0$ considered so far are much larger than
the value $\Omega_{0}\approx 0.2$ obtained from most observations when
luminous matter and the dark matter inferred from galaxy motions are
added (see e.g. Riess et al. \cite{riess}), and they can be only
explained by assuming large amounts of as yet unobserved dark matter.
In the standard model the observational value $\Omega_{0}\approx 0.2$
can be only obtained for $k=-1$ while in the present model
$\lambda_{0}>0.8$ is required.  With these choices the differences
between the two models are remarkable: In the standard model the
expansion is decelerated ($q_{0}=0.1$) while it is accelerated in the
present model ($q_{0}\le -0.7$), and in the present model the scale
factor $S_{0}$ of today and the life time $t_{0}$ of the universe
after the phase transition, that can be calculated from (\ref{15}) and
(\ref{12}), are much larger than in the standard model.  For
convenience the values of $S_{0}$, $q_{0}$ and $t_{0}$ obtained for
the standard model, the present model and other models with
$\lambda_{0}>0$ are listed in Table~\ref{tab:1} for different choices
of the parameters $\lambda_{0}$, $\Omega_{0}$ and $k$. For the
evaluation of $S_{0}$ and $t_{0}$ the value $H_{0}=65$
km\,s$^{-1}$\,Mpc$^{-1}$ of the Hubble parameter, having a high probability
according to latest measurements (see Riess et al. \cite{riess}), was
used.

\begin{table}
\begin{tabular}{lllllll}
\hline\hline
Model& $\lambda_{0}$ & $\Omega_{0}$& $k$ & $S_{0}$ & $q_{0}$ & $t_{0}$ \\
\hline 
standard      &0    &0.2   &-1               &16.9 &\hphantom{-}0.1  &12.8\\
standard      &0    &1     &\hphantom{-}0    &---  &\hphantom{-}0.5  &\hphantom{0}8.7\\
stand./pres.  &0    &1.1   &\hphantom{-}1    &41.3 &\hphantom{-}0.55 &\hphantom{0}8.5\\
\hline 
present       &0.85 &0.2   &\hphantom{-}1    &67.4 &-0.75            &16.6\\
\hline
other         &0.75 &0.2   &-1               &67.4 &-0.65            &15.9\\ 
other         &0.8 &0.2    &\hphantom{-}0    &---  &-0.7             &16.2\\ 
\hline\hline
\end{tabular}
\caption{\label{tab:1}Typical parameter values obtained for the 
  standard model, the present model and other models ($\lambda_{0}>0$
  and $k<1$) with $H_{0}=65$ km\,s$^{-1}$\,Mpc$^{-1}$.
  $S_{0}$ is given in $10^{9}$ light-years and the life time $t_{0}$ of
  the universe after the phase transition is given in $10^{9}$ years.
  (For $k=0$ no value of $S_{0}$ follows from (\ref{15}).)}
\end{table}

The closure of the universe, associated with $k=1$, could in principle be
detected, e.g. by the double observation of a very bright and distant
object in opposite directions. So far searches for double observations
have not been successful, but if the present model would apply, the
result $S_{0}=67.4\;\mathrm{ly}\gg ct_{0}=16.6\;\mathrm{ly}$ would
explain why.

It can be concluded in summary that pronounced differences between the
present model and standard big bang models with inflation only appear
if the universe does not contain large amounts of dark matter, and
thus $\Omega_{0}$ is markedly smaller than $1$. The evaluation of
observational data some time ago (Liebscher et al. \cite{liebscher})
and just recently (Riess et al. \cite{riess}, Perlmutter et al.
\cite{perlmutter} and Branch \cite{branch}) yields a strong preference
for a non-negligible positive cosmological constant of about the
magnitude that was employed for the evaluation of the present model in
the case $\Omega_{0}=0.2$. The recent observational data also confirm
the negative value of the deceleration parameter $q_{0}$ connected
with it. It can be seen from Table~\ref{tab:1} that for comparable
values of $\Omega_{0}$ and $\lambda_{0}$ in an open universe ($k=0$
and $k=-1$) almost the same results are obtained.

It is illuminating to check the ``strong energy condition''
\begin{displaymath}
  \varrho_{\mathrm{m}} + \frac{3 p}{c^{2}}-\frac{\Lambda c^{2}}{4\pi G}>0
\end{displaymath}
that must be satisfied for all times by solutions starting with a big
bang singularity (see e.g. Wald \cite{wald}). With the assumptions
underlying the present model the condition becomes
\begin{displaymath}
  2(\varrho_{\mathrm{m}} - \varrho_{\mathrm{v}})>0\,.
\end{displaymath}
The equilibrium from which the present model evolves through
instability has $\varrho_{\mathrm{m}} = \varrho_{\mathrm{v}}$ and thus
marks the boundary of the regime in which the strong energy condition
is not satisfied; in the later evolution $\varrho_{\mathrm{m}}$
decreases while $\varrho_{\mathrm{v}}$ remains constant so the left
hand side of the inequality becomes negative and thus recedes from the
boundary.

It is possible to attribute the vacuum energy density of the present
model to the same kind of scalar field as introduced by Starkovich
\& Cooperstock (\cite{starkovich}) or by Bayin et al.  (\cite{bayin}).
In this case the treatment is only slightly modified, and the main
results are essentially the same as obtained in this section as will
be shown in Sect.~\ref{sec:vacen}.

\section{Classification of inflationary solutions with $\Lambda=$
  const and $k=1$}

In this section it is shown how the inflationary branch of the soft
bang model presented in this paper fits into the framework of general
inflationary solutions with $\Lambda$ = const (or
$\varrho_{\mathrm{v}}=$ const equivalently) for $k=1$. With $\tau=ct$,
$\;\alpha=8\pi G/(3c^{2})$ and $\;C=\varrho_{\mathrm{m}}S^{4}$ Eq.
(\ref{5}) becomes
\begin{equation}\label{17}
  \dot{S}^{2}(\tau)+V(S)=0
\end{equation}
with
\begin{equation}\label{18}
 V(S)=1-\frac{\alpha C}{S^{2}}-\alpha\varrho_{\mathrm{v}}S^{2}\,. 
\end{equation}
1. \emph{Case} $C=0$\\
This is the matter free case of the IR-model  or the
BP-model resp. According to (\ref{17}), $S(t)$ is restricted to values
$S\ge S_{*}$ with $V(S_{*})=0$, since \mbox{$V(S)=-\dot{S}^{2}\le 0$.}
From (\ref{18}) one obtains
\begin{equation}\label{19}
  S_{*}=\frac{1}{\sqrt{\alpha\varrho_{\mathrm{v}}}}
\end{equation}
with $\dot{S}=0$ for $S=S_{*}$ according to (\ref{17}).  In the
solution of the BP-model, $S(t)$ comes from $\infty$ for $t\to
-\infty$, decreases until a minimum value
$S_{\mathrm{min}}\gg l_{\mathrm{Pl}}$ is reached, and then turns around
to increasing values.  This solution is obtained from (\ref{17}) for
$S_{*}=S_{\mathrm{min}}\gg l_{\mathrm{Pl}}$.

On the other hand, the solution of the IR-model is obtained for
$S_{*}<l_{\mathrm{Pl}}$, starting at $S=l_{\mathrm{Pl}}$ with
$\dot{S}>0$ and $\ddot{S}>0$.

\begin{figure}[tb]
  \includegraphics[width=0.5\columnwidth]{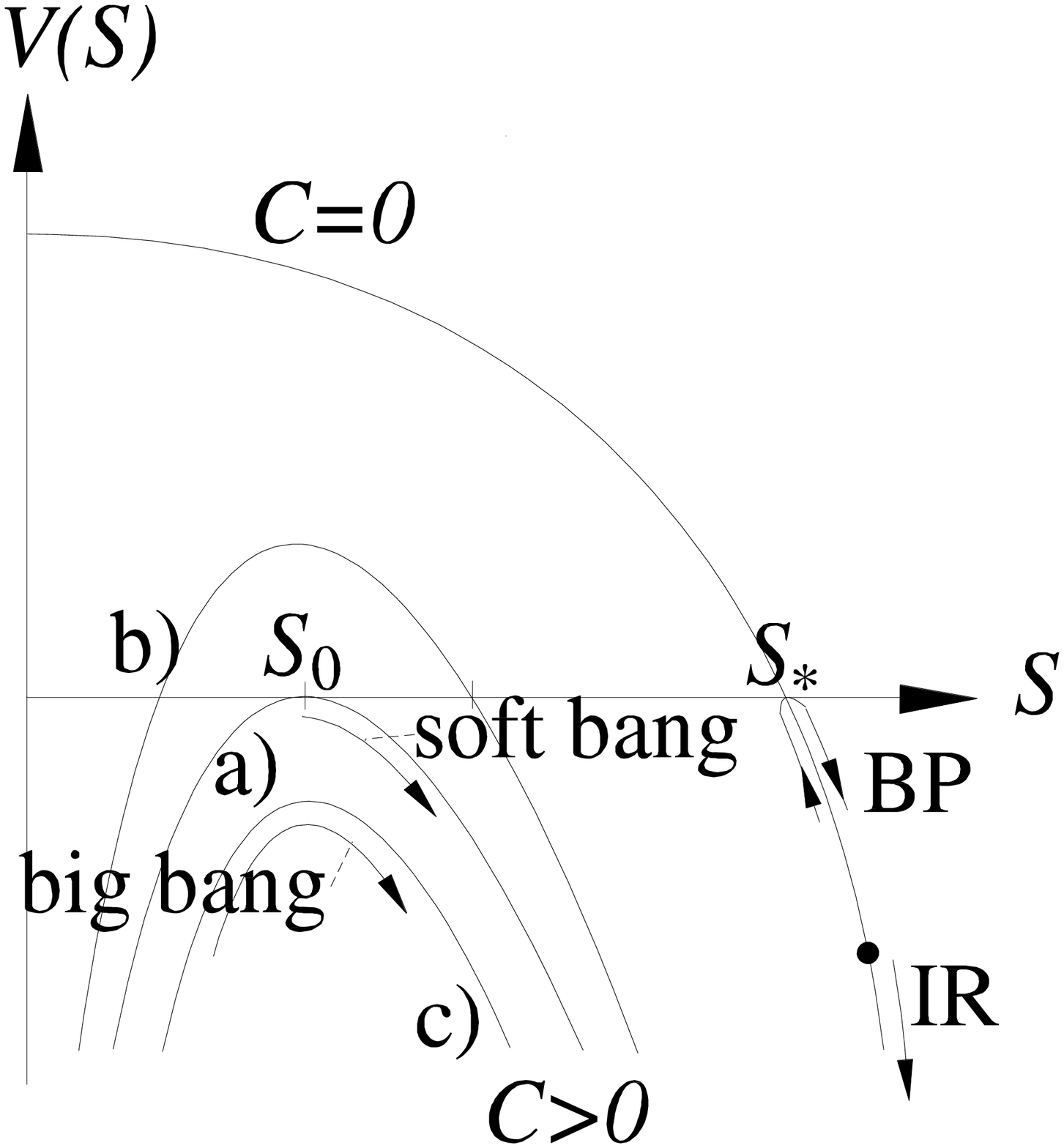}\\
  \caption{\label{pic:2}Potential $V(S)$ for the cases $C=0$ and 
    $C>0$. In the case $C>0$, curve a) applies
    for$V_{\mathrm{max}}=0$, b) for $V_{\mathrm{max}}>0$ and c) for
    $V_{\mathrm{max}}<0$.}
\end{figure}

According to (\ref{19}) $S_{*}=l_{\mathrm{Pl}}$ for
\begin{equation}\label{20}
  \varrho_{\mathrm{v}}=\frac{1}{\alpha l^{2}_{\mathrm{Pl}}}
  =:\varrho_{\mathrm{v}*}\,,
\end{equation}
and thus the following classification is obtained:
\begin{eqnarray*}
  \mbox{solution~ of~ the BP-model}\qquad&\mbox{for}&\qquad
  \varrho_{\mathrm{v}}<\varrho_{\mathrm{v}*}\\*
  \mbox{solution~ of~ the IR-model}\qquad&\mbox{for}&\qquad 
  \varrho_{\mathrm{v}}>\varrho_{\mathrm{v}*}\,.
\end{eqnarray*}
2. \emph{Case} $C>0$\\
Fig.~\ref{pic:2} shows the potential $V(S)$ with $C>0$ for three
different kinds of solution together with its shape for $C=0$. For
$C>0$ it has a maximum
\begin{equation}\label{21}
  V_{\mathrm{max}}=1-2\alpha\sqrt{C\varrho_{\mathrm{v}}}
\qquad\mbox{at}\qquad
S_{0}=\left(\frac{C}{\varrho_{\mathrm{v}}}\right)^{1/4}\,.
\end{equation}

\begin{figure}[tb]
  \includegraphics[width=0.7\columnwidth]{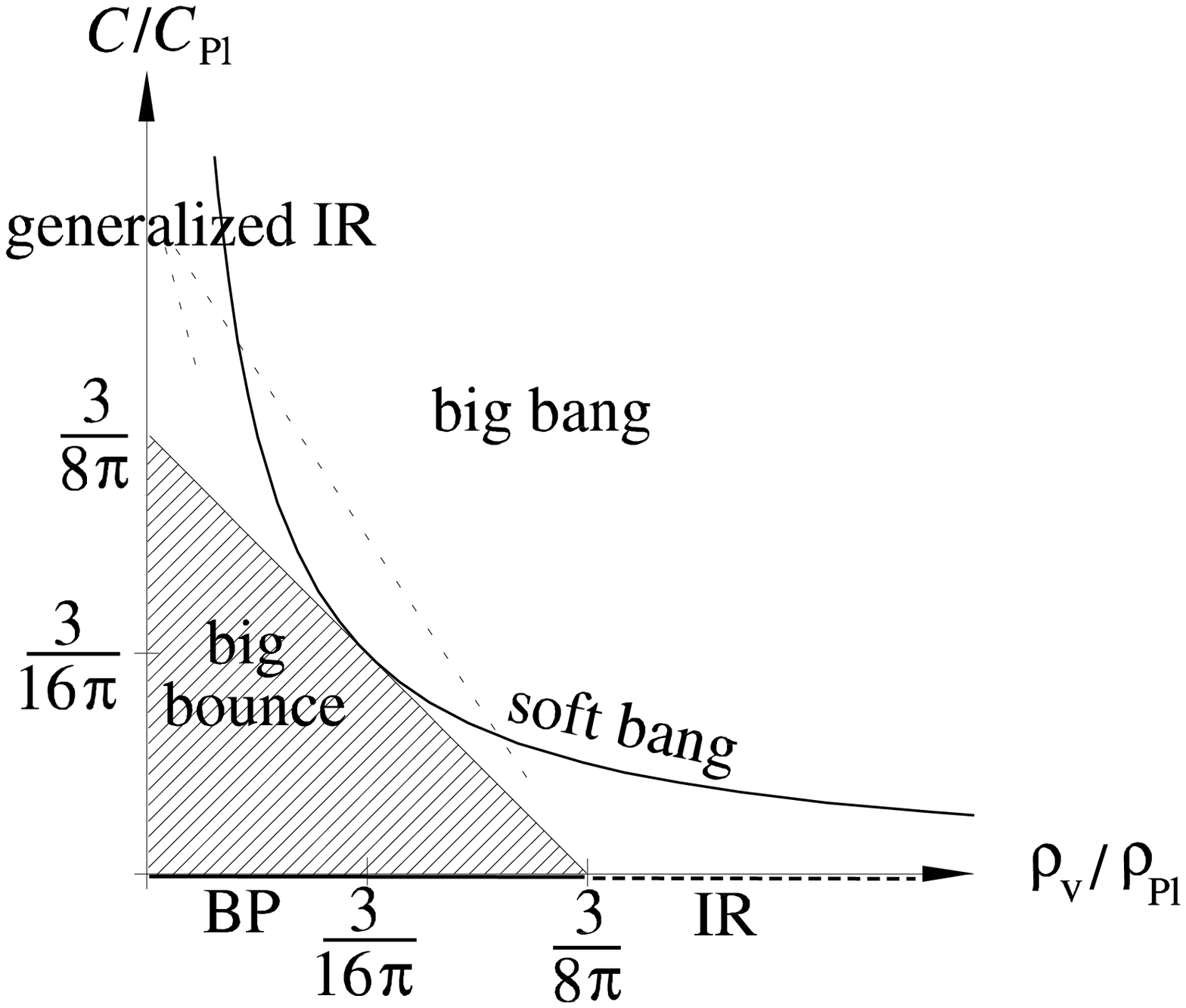}\\
  \caption{\label{pic:3}Diagram $C/C_{\mathrm{Pl}}$ versus 
    $\varrho_{\mathrm{v}}/\varrho_{\mathrm{Pl}}$ with location of the
    different kinds of inflation solutions with $k=1$. The curve
    \emph{soft bang} corresponds to (\ref{22}) and is the
    location of soft bang solutions, the upper bound of the shaded
    area corresponds to (\ref{24}). BP-model solutions are located on
    the $\varrho_{\mathrm{v}}/\varrho_{\mathrm{Pl}}$ axis in the range
    from $0$~to~$3/(8\pi)$, IR-model solutions in the range
    $\varrho_{\mathrm{v}}/\varrho_{\mathrm{Pl}}>3/(8\pi)$. In the shaded area
    generalized big bounce solutions are obtained, in the area above
    it and below the soft bang curve generalized IR-solutions. The
    region above the soft bang curve is the location of big bang
    solutions.}
\end{figure}

\noindent
a) For $V_{\mathrm{max}}=0$ or
\begin{equation}\label{22}
  C=\frac{1}{4\alpha^{2}\varrho_{\mathrm{v}}}\,,
\end{equation}
there is an unstable equilibrium point at $S=S_{0}$, and the soft bang
solution of this paper is obtained. The curve given by (\ref{22}) is
shown in Fig.~\ref{pic:3}.

\noindent
b) For $V_{\mathrm{max}}>0$ or
$C<1/(4\alpha^{2}\varrho_{\mathrm{v}})$, big bounce solutions of the
same type as in the BP-model are possible as well as solutions of the
type used in the IR-model, the difference being that a matter and/or
radiation density $\varrho_{\mathrm{m}}=C/S^{4}\ne 0$ coexists with
the vacuum energy density $\varrho_{\mathrm{v}}$.  Since for all
solutions \mbox{$S_{0}=(C/\varrho_{\mathrm{v}})^{1/4}$}, it follows
that $C=\varrho_{\mathrm{v}}S^{4}_{0}=\varrho_{\mathrm{m}}S^{4}$.
Now, $S(t)$ is restricted to $S\ge S_{*}>S_{0}$ (see Fig.~\ref{pic:2})
with
\begin{equation}\label{23}
  S_{*}=\frac{1}{\sqrt{2\alpha\varrho_{\mathrm{v}}}}\,
  \left(1+\sqrt{1-4\alpha^{2}\varrho_{\mathrm{v}}C}\right)^{1/2}
\end{equation}
obtained from $V(S_{*})=0$, and therefore
\begin{displaymath}
  \varrho_{\mathrm{m}}=\varrho_{\mathrm{v}}(S_{0}/S)^{4}\le
  \varrho_{\mathrm{v}}(S_{0}/S_{*})^{4}<\varrho_{\mathrm{v}}\,.
\end{displaymath}
Generalized big bounce solutions of the type considered in the
BP-model are obtained for \mbox{$S_{*}>l_{\mathrm{Pl}}$}, and
generalized solutions of the IR-model type for
$S_{*}<l_{\mathrm{Pl}}$. The boundary between the two is given by
$S_{*}=l_{\mathrm{Pl}}$ which with (\ref{23}) leads to the boundary
equation
\begin{equation}\label{24}
  C=l^{2}_{\mathrm{Pl}}\left(\frac{1}{\alpha}-
    \varrho_{\mathrm{v}}l^{2}_{\mathrm{Pl}}\right)
\end{equation}
represented in Fig.~\ref{pic:3} by the upper boundary of the
shaded area.

\noindent
c) For $V_{\mathrm{max}}<0$ or $C>1/(4\alpha^{2}\varrho_{\mathrm{v}})$
finally,  big bang solutions with $k=1$ are obtained with 
$\varrho_{\mathrm{m}}=C/S^{4}\to\infty$ for $S\to 0$. 

Fig.~\ref{pic:3} shows where the different kinds of inflation
solutions with $k=1$ are located in a $C/C_{\mathrm{Pl}}$ versus
$\varrho_{\mathrm{v}}/\varrho_{\mathrm{Pl}}$ diagram, the definitions
$l_{\mathrm{Pl}}=(\hbar G/c^{3})^{1/2}$,
$\varrho_{\mathrm{Pl}}=c^{5}/(\hbar G^{2})$ and
\mbox{$C_{\mathrm{Pl}}=\varrho_{\mathrm{Pl}}l^{4}_{\mathrm{Pl}}$}
being used.  The boundary curves (\ref{20}) and (\ref{24}) don't
intersect but only touch at
\begin{displaymath}
  \varrho_{\mathrm{v}}=\frac{1}{2\alpha l^{2}_{\mathrm{Pl}}}
  =\frac{3}{16\pi}\varrho_{\mathrm{Pl}}\,,
  \qquad
  C=\frac{l^{2}_{\mathrm{Pl}}}{2\alpha}=\frac{3}{16\pi}C_{\mathrm{Pl}}\,.
\end{displaymath}

\section{Attribution of the vacuum energy to a scalar quantum field}
\label{sec:vacen}

In this section it is shown that the vacuum energy can be attributed
to a scalar quantum field $\Phi$ in essentially the same way as by
Starkovich \& Cooperstock (\cite{starkovich}) and by Bayin et al.
(\cite{bayin}), only some slight modifications being necessary. For
clearness the essential steps of their approach is briefly
recapitulated.

The scalar field $\Phi$ that is responsible for the vacuum energy
density is conformally coupled to the Ricci-curvature $R$ and is
described by a generalized Klein-Gordon equation
\begin{equation}\label{25}
  g^{\mu\nu}\partial_{\mu}\partial_{\nu}\Phi +\xi R\Phi +dV(\Phi)/d\Phi =0
\end{equation}
where $V(\Phi)$ is the scalar field potential and $\xi$ a numerical
constant. According to Birrel \& Davies (\cite{birrell}) the
energy-momentum-tensor of the field is
\begin{eqnarray}
  T_{\mu\nu}&=&(\varrho_{\mathrm{v}} +
  p_{\mathrm{v}}/c^{2})u_{\mu}u_{\nu}-p_{\mathrm{v}}
  g_{\mu\nu}\qquad\mbox{with}\label{26}\\
  u_{\mu}&=&\frac{\partial_{\mu}\Phi}{\left(\partial_{\alpha}\Phi
      \partial^{\alpha}\Phi\right)^{1/2}}\,,\label{27}\\
  \varrho_{\mathrm{v}} c^{2}&=&\frac{\dot{\Phi}^{2}}{2}+\frac{\Phi^{2}}{2}
  \left[\left(\frac{\dot{S}}{S}\right)^{2}+\frac{1}{S^{2}}\right]
  +\frac{\dot{S}\Phi\dot{\Phi}}{S}+V(\Phi)\,,\label{28}\\
  p_{\mathrm{v}}&=&\frac{\dot{\Phi}^{2}}{6}+\frac{\Phi^{2}}{6}
  \left[\left(\frac{\dot{S}}{S}\right)^{2}+\frac{1}{S^{2}}\right]
  +\frac{\dot{S}\Phi\dot{\Phi}}{3S}-V(\Phi)\\
  &&+\frac{\Phi\,dV(\Phi)/d\Phi}{3}\,.
\label{29}
\end{eqnarray}
With this from Einsteins field equations for $k=1$ the equations
\begin{equation}\label{30}
  \dot{S}^{2} 
  = \frac{8\pi G}{3} \varrho_{\mathrm{v}} S^{2}- c^{2} \,,
\end{equation}
\begin{equation}\label{31}
  \ddot{S} = \frac{8\pi G}{3} \left(1-\frac{3\gamma}{2}\right)
  \varrho_{\mathrm{v}}\,S
\end{equation}
are obtained where
\begin{equation}\label{32}
  \gamma:=1+\frac{p_{\mathrm{v}}}{\varrho_{\mathrm{v}} c^{2}}\,.
\end{equation}
Differentiating (\ref{30}) with respect to time $t$ yields
\begin{displaymath}
  \frac{d\varrho_{\mathrm{v}}}{dS}=\frac{\dot{\varrho}_{\mathrm{v}}}{\dot{S}}
  = \frac{3}{4\pi GS^{2}}\left(\ddot{S}-\frac{8\pi G}{3} 
    \varrho_{\mathrm{v}}S\right)\,,
\end{displaymath}
and eliminating $\ddot{S}$ from this equation with (\ref{31}) yields
$d\varrho_{\mathrm{v}}/dS=-3\gamma\varrho_{\mathrm{v}}/S$ and
\begin{equation}\label{33}
 \varrho_{\mathrm{v}}=\frac{\varrho_{*}S^{3\gamma}_{*}}{S^{3\gamma}}\,. 
\end{equation}
From (\ref{30}) and (\ref{31}) $\varrho_{\mathrm{v}}$ can be
eliminated yielding 
\begin{displaymath}
  \ddot{S}+\left(\frac{3\gamma}{2}-1\right)
  \left(\frac{\dot{S}^{2}+c^{2}}{S}\right)=0\,.
\end{displaymath}
From this equation the time evolution of $S$ can be determined
independently of the evolution of the field $\Phi$ if $\gamma$ is
prescribed. Bayin et al.  (\cite{bayin}) made the simplifying
assumption that during different eras in the evolution of the universe
the quantity $\gamma$ defined in (\ref{32}) assumed different but
constant values, especially a very small one, $\gamma\approx 10^{-3}$,
in the inflationary era.

With slight modifications these ideas can be incorporated into the
present model. For this purpose it is assumed, that the field $\Phi$
can be present in addition to primordial matter of equal density
$\varrho_{\mathrm{m}}$ and has the same properties as without matter.
(In order to explain the primordial equipartition between energy of
the field $\Phi$ and energy of matter, some interaction should be
present, but this has to be so small that it can be neglected for the
dynamical evolution.)

With this assumption, Eqs. (\ref{30})--(\ref{31}) for $k=1$ must be
replaced by the equations
\begin{eqnarray}
  \dot{S}^{2}(t) &=& \frac{8\pi G}{3}\,
  (\varrho_{\mathrm{m}} + \varrho_{\mathrm{v}}) S^{2} - c^{2}\,, \label{5*}\\
 \ddot{S}(t) &=&- \frac{8\pi G}{3}\left[\varrho_{\mathrm{m}} 
 - \left(1-\frac{3\gamma}{2}\right)\varrho_{\mathrm{v}}\right]\,S\,,\label{6*}
\end{eqnarray}
and Eq. (\ref{33}) must be employed instead of the former assumption
$\varrho_{\mathrm{v}}=\mathrm{const}$ that is recovered from
(\ref{33}) for $\gamma=0$. 

For an equilibrium $\dot{S}\equiv0$ and $\ddot{S}\equiv0$ must be
satisfied, and in order to obtain the same equilibrium as in
Sect.~\ref{sec:softb} the assumption $\gamma=0$ must be made. With
this, from (\ref{5*})--(\ref{6*}) one obtains
$\varrho=\varrho_{\mathrm{v}}=:\varrho_{*}$ and
\begin{equation}\label{34}
  S=S_{*}=c\sqrt{\frac{3}{16\pi G \varrho_{*}}}
\end{equation}
as equilibrium conditions. The dynamics of deviations from equilibrium
is obtained from from (\ref{30}) with (\ref{4}) and (\ref{33}), and
similarly as (\ref{9}) one now obtains the equation
\begin{equation}  \label{35}
  \dot{S}^{2}=\frac{c^{2}}{2}
  \left(\frac{S^{2}_{*}}{S^{2}}+\frac{S^{2-3\gamma}}{S^{2-3\gamma}_{*}}-2\right)
\end{equation}
for it. With $\gamma =0$ the same results as in Sect.~\ref{sec:softb}
would be obtained. An evolution similar to that obtained by Bayin et
al. (\cite{bayin}) can be achieved by assuming that $\gamma$ increases
as soon as the equilibrium is left, and saturates at the small value
\mbox{$\gamma\approx 10^{-3}$} in order to obtain inflation.

For studying the separation of $S(t)$ from the equilibrium value
$S_{*}$, the ansatzes
\begin{displaymath}
  S=S_{*}+s=S_{*}(1+\epsilon)
\quad\mbox{with}\quad
\epsilon=s/S_{*}
\end{displaymath}
and
\begin{displaymath}
  \gamma=\alpha \epsilon
\end{displaymath}
with some constant $\alpha$ are made. Expansion of (\ref{35}) with
respect to $\epsilon$ up to terms of order $\epsilon^{2}$ yields
\begin{displaymath}
 \dot{S}^{2}=\frac{c^{2}}{2}(4-3\alpha)\epsilon^{2} \,,
\end{displaymath}
and for real solutions $\alpha<4/3$ must be assumed in addition.  With
these assumptions an expansion evolution of the universe from an
initial equilibrium state through instability becomes possible as in
Sect.~\ref{sec:softb}.

For $S\gg S_{*}$ with the assumption $\gamma=\mathrm{const}\approx
10^{-3}$ Eq.~(\ref{35}) can be approximated by
\begin{displaymath}
  \dot{S}^{2}=\frac{c^{2}}{2}
  \left(\frac{S^{2-3\gamma}}{S^{2-3\gamma}_{*}}\right)\,,
  \qquad
  \dot{S}=\frac{c}{\sqrt{2}}\left(\frac{S}{S_{*}}\right)^{1-3\gamma/2}
\end{displaymath}
and yields an evolution
\begin{displaymath}
  S\sim (t-t_{0})^{2/(3\gamma)}\,.
\end{displaymath}
Since $2/(3\gamma)$ is very large, this is an inflation-like evolution
although algebraic instead of exponential. At some stage and with
similar assumptions and consequences as in Sect.~\ref{sec:softb} a
phase transition described by a rapid change of $\gamma$ must take
place in order to enable the transition to a Friedmann-Lema{\^\i}tre
evolution.

In the primordial equilibrium state $\gamma=0$ viz.
$p_{\mathrm{v}}=-\varrho_{\mathrm{v}} c^{2}$, $S=S_{*}$ and
$\dot{\Phi}=0$. With this from (\ref{28})--(\ref{29}) the condition
\begin{equation}\label{36}
  \frac{dV(\Phi)}{d\Phi}=-\frac{2\Phi_{*}}{S^{2}_{*}}
\end{equation}
is obtained, and with the choice $\Phi_{*}=0$ it can be achieved that
$V(\Phi)$ has an extremum. $V(\Phi)$ is an arbitrary function and
since no condition is obtained for $V(\Phi_{*})=V(0)$, it appears that
this quantity, which is contained in (\ref{28})--(\ref{29}) and
(\ref{31}), can be chosen such that the extremum becomes a minimum.

\section{Observational constraints}
\label{sec:obs-cons}

Models with a cosmological term $\Lambda$ (or $\varrho_{\mathrm{v}}$
equivalently) must comply with certain observational constraints for
which a clear survey was given by Overduin \& Cooperstock
(\cite{overduin}).

Of course, the flatness constraint $\Omega_{0}+\lambda_{0}=1$, where
$\Omega_{0}$ and $\lambda_{0}$ are the present values of
$\Omega=\varrho_{\mathrm{m}}/\varrho_{\mathrm{cr}}$ and
$\lambda=\varrho_{\mathrm{v}}/\varrho_{\mathrm{cr}}$ with
$\varrho_{\mathrm{cr}}=3H^{2}/(8\pi G)$, cannot apply for the present
model since a space-time with positive curvature can never become
completely flat.  

Observations concerning CBM fluctuations, gravitational lens
statistics, supernovas etc. restrict $\lambda_{0}$ to a range
$0.5-0.8$. The value $0.85$ used for the calculations in this paper is
in fair agreement with this and can be taken even smaller when
$\Omega_{0}$ is raised correspondingly.

The age of the universe is another quantity imposing rather stringent
conditions on possible values of $\lambda_{0}$. The present model has
an infinite age of the universe. However, the time $t_{0}$ elapsed
after the phase transition until today, coinciding with the time that
was available for the creation of the elements observed in the
universe and the evolution of stars and galaxies, should observe the
same conditions as the age of universes with finite past. It was
numerically evaluated for the present model, and its value presented
in Table~\ref{tab:1} is in very good agreement with the latest
requirements derived from observational data (see e.g. Riess et al.
\cite{riess}).

Another constraint is provided by the requirement that, in a closed
universe, the antipode must be further away than the most distant
object for which gravitational lensing is observed. For the present
model, according to Table~\ref{tab:1} the distance of our antipode is
$\pi S_{0}= \pi\cdot 67{.}4\cdot 10^{9}$~ly which is still far beyond
our horizon, so the gravitational lensing constraint is well observed.

By nonsingular models the maximum red-shift constraint must be observed
which requires that the maximal value of the red-shift
\begin{displaymath}
  z=\frac{S(t_{0})}{S(t_{\mathrm{em}})}-1
\end{displaymath}
($t_{\mathrm{em}}=$ time of emission), obtained by inserting for
$S(t_{\mathrm{em}})$ the smallest value that $S$ assumes in the model,
must be at least as large as the greatest red-shift observed.  In the
present model, $S(t_{0})=S_{0}=67{.4}\cdot 10^{9}$ ly, the minimal
value of $S$ is \mbox{$S_{*}\approx 3\cdot 10^{-27}\,\mathrm{m}$}, and
therefore the greatest possible red-shift is much greater than the
greatest one observed.

\section{Discussion and summary}

In the present model all kinds of singularities (e.g. of the expansion
velocity, the expansion rate, or $S\to \infty$ for $t\to -\infty$ as
in the BP-model) are avoided and no necessity arises for employing a
theory of quantum gravity. For $t\to -\infty$ the (closed) universe is
a tiny micro-universe in a classical static state.  Its
expansion is triggered by an instability and starts quite slowly at
the velocity $\dot{S}=0$ and expansion rate $\dot{S}/S=0$.  Only much
later it gains appreciably until it becomes exponentially inflating
and finally reaches the later expansion rate of an inflationary big
bang universe. It is an advantage of the present model that
the universe is not ``born'' with an unexplained and extreme expansion
rate like in most big bang models, but that the observed expansion can
be explained as the consequence of an instability in its far past.

It should be noted that the dynamical solution describing the
departure from the unstable equilibrium obtained in this paper is
quite different from the well known Eddington-Lema{\^\i}tre solution
of the Einstein-Lema{\^\i}tre equations that describes the departure
from Einstein's static solution. While it has been shown by
B{\"o}rner \& Ehlers (\cite{boerner}) and Ehlers \& Rindler (\cite{ehlers})
that for $\Omega_{0}\ge 0.02$ the latter violates the maximum
red-shift constraint, the present model is in good agreement with this
constraint and other observations, as was shown in
Sect.~\ref{sec:obs-cons}.

The coexistence of a quantum field of energy density
$\varrho_{\mathrm{v}}c^{2}$ with some sort of primordial relativistic
matter and/or radiation is an essential ingredient of the present
model that may be critically considered and certainly needs
discussion. The fact that a similar coexistence is assumed in big bang
models with inflation may be invoked in support, but it may be a weak
argument in view of opposing arguments raised by other authors.
Priester et al. (\cite{priester}) emphasize that a quantum vacuum state
of the universe is the more natural choice for its primordial stage
than a state in which elementary particles already have been present.
Usually the origin of the vacuum energy density is assumed to be
either the Higgs field that was introduced in elementary particle
physics in order to explain the mass of elementary particles through
interactions, or some other quantum field or other causes like
worm-holes etc. The question is whether a primordial quantum field can
exist on its own as a precondition for massive particles to be formed
later, or whether it in turn needs these particles for its own
existence. When the venerable principle \emph{actio = reactio} is
invoked, the second view appears as the more natural one. If this is
accepted, then still another feature of the model becomes plausible.
At first glance an extreme fine tuning appears to be needed in order
that the condition $\varrho_{\mathrm{m}}=\varrho_{\mathrm{v}}$ is
getting satisfied.  However, in an equilibrium state equipartition
between two interacting ingredients is the only natural constellation,
and in this spirit the fact, that equipartition in the equilibrium
state results from the the field equations, may even appear as a
confirmation.

It is interesting to note that the condition for the existence of an
unstable equilibrium state from which the universe evolves through
instability is quite contrary for the classical equilibrium of the
present model and the quantum equilibrium considered by Starobinsky
(\cite{starobinsky}): in the classical case in addition to the Higgs
field the presence of particles is necessary while in the quantum case
just the absence of particles is required.

In this paper, in agreement with its modern interpretation as the
energy density of some quantum field, the cosmological constant is
treated as a dynamical variable rather than a geometrical one.
However, this treatment is rather crude because $\Lambda$ is still
kept constant for most of the time -- at a very large value during
early stages of the universe and at its present low value after the
phase transition until today. Many models have been proposed coupling
the decay of $\varrho_{\mathrm{v}}$ to the time evolution of the
universe or its scale factor $S$ - a survey is presented in a paper by
Overduin \& Cooperstock (\cite{overduin}) -, and it should be possible
to replace the instant transition assumed in this paper with one of
these more refined transitions without major changes in the results.
The validity of this assumption was demonstrated in
Sect.~\ref{sec:vacen} for one specific model.

\begin{figure}[tb]
  \includegraphics[width=0.7\columnwidth]{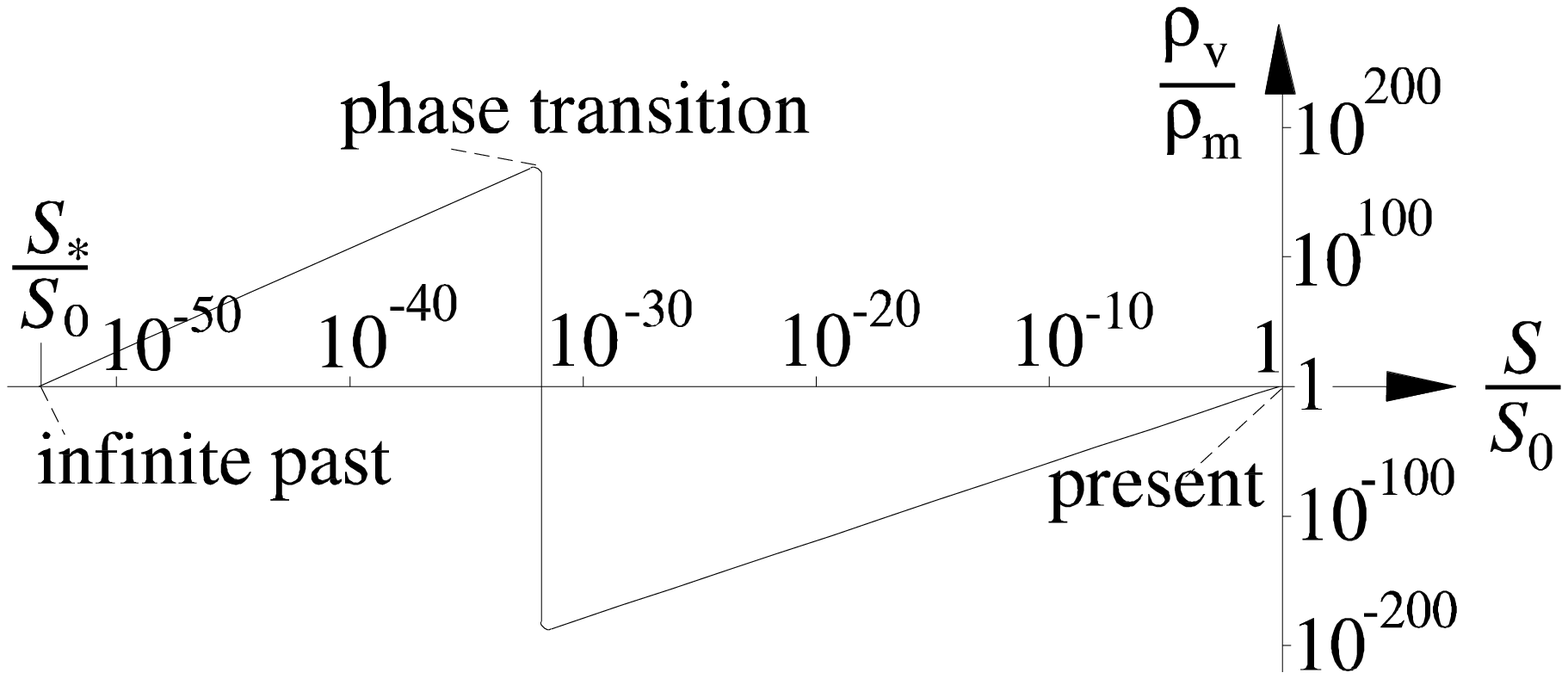}\\
  \caption{\label{pic:4}. Ratio
    $\varrho_{\mathrm{v}}/\varrho_{\mathrm{m}}$ as function of $S/S_{0}$
  for the present model.}
\end{figure}

A look at the evolution of $\varrho_{\mathrm{v}}/\varrho_{\mathrm{m}}$
used for the calculations of this paper (Fig.~\ref{pic:4}) shows
that the present model could contribute to a solution of the
``coincidence problem''. This is raised by the latest observations
according to which $\varrho_{\mathrm{v}}$ lies in the same range as
the matter density $\varrho_{\mathrm{m}}$ today (see Zlatev et al.
\cite{zlatev}). The coincidence problem consists in the fact that,
according to many models, $\varrho_{\mathrm{m}}$ and
$\varrho_{\mathrm{v}}$ start at very different values in the early
universe and require an extreme fine tuning at that time in order to
reach almost equal values at present.  In the soft bang model of this
paper the universe starts with
$\varrho_{\mathrm{m}}=\varrho_{\mathrm{v}}$, during the evolution of
the universe there are times at which the two densities temporarily
depart from each other, but today they are very close to each other
again. It appears that the present model would provide a good starting
point for developing a quintessence field with ``tracking properties''
(see Zlatev et al. \cite{zlatev}) -- at least it appears to be in good
agreement with the requirements of such a concept.

Conceptually it has been considered as a very satisfying property of
big bang models that in them the universe does not have an eternal
past but originated from some act of creation. In this sense the
eternal past of the present model may appear as a conceptual
disadvantage.  However, the situation is not as bad as it appears. In
physics time is a parameter that is used for ordering changes of
states. However, when there are no changes then this order parameter
loses its sense. In a very slowly changing situation it may therefore
become more useful to consider the changes themselves as the order
parameter that represents time instead of using an order parameter
ordering no changes. In this sense the lifetime of the universe
considered in this paper is not greater than that of big bang models
because there has been even a smaller change from the original state
to the present.

\end{document}